\documentclass[fleqn,usenatbib]{mnras}
\usepackage{float}
\usepackage[T1]{fontenc}
\usepackage{ae,aecompl}
\usepackage{graphicx}
\usepackage{epstopdf}
\usepackage{amsmath}
\usepackage{amssymb}
\usepackage{supertabular}
\usepackage{siunitx}
\usepackage{caption}

\def\beq{\begin{eqnarray}}
\def\eeq{\end{eqnarray}}

\title[RFRBs from IMBH-WD TDEs]{Short-lived repeating fast radio bursts from tidal disruption of white dwarfs by intermediate-mass black holes}
\author[Xing \& Liu]{Jing-Tong Xing, Tong Liu\thanks{E-mail: tongliu@xmu.edu.cn}\\
Department of Astronomy, Xiamen University, Xiamen, Fujian 361005, China}

\begin{document}
\label{firstpage}
\pagerange{\pageref{firstpage}--\pageref{lastpage}}
\maketitle
	
\begin{abstract}
The origin of repeating fast radio bursts (RFRBs) is still a mystery. We propose that short-lived RFRBs might be triggered from the tidal disruption of white dwarfs (WDs) by intermediate-mass black holes (BHs). In this model, we show that the remnant WD clusters after tidal collapse cuts the magnetic lines on the BH accretion discs, and during each fall of the clump, so that electrons are torn from the surface of the mass and instantly accelerated to the relativistic energy. The subsequent movement of these electrons along magnetic field lines will result in coherent curvature radiation. This short-lived radio transients might accompany with the accretion process. The luminosity and the timescale can be estimated to be $L_\mathrm{tot}\sim 1.96\times10^{40}~{\rm erg~s^{-1}}$ and $\Delta t\sim1.14~{\rm ms}$, respectively, which are consistent with the typical properties of RFRBs. Moreover, the total event rate of our model for generating RFRBs might be as high as $\sim 10~\rm {yr^{-1}~Gpc^{-3}}$.
\end{abstract}
	
	
\section{Introduction} \label{sec:introduction}

Fast radio bursts (FRBs) are millisecond radio transient sources, dating back as early as 2007 when an unknown bright single pulse was detected in archival data from the Parkes Telescope \citep{2007Sci...318..777L}, marking the beginning of research in this area. At present, with the continuous improvement of the accuracy of observation instruments, the observed data of FRBs are also increasing \citep[e.g.][]{2020Natur.587...63L,2021ApJS..257...59C,2021Natur.598..267L}, but the origin of FRBs has not been solved. According to the observational properties, FRBs can be divided into two types: repeating FRBs (RFRBs) and one-off FRBs. The two types of FRBs differ in the interpretation models of origin. To model RFRBs, the most important thing is to physically explain the typical duration and the burst luminosity of FRBs. Available evidences suggest that FRB core engines are compact, and that the high luminosity of radio signals requires coherent radiation \citep[e.g.][]{2014PhRvD..89j3009K,2014ApJ...785L..26L}. These requirements have led some authors to associate compact objects such as white dwarfs (WDs), neutron stars (NSs), and black holes (BHs). Relevant models include but not limited to the young, rapidly rotating pulsars \citep[e.g.][]{2016MNRAS.462..941L}, the pulsars that produce huge pulses \citep[e.g.][]{2016MNRAS.458L..19C,2016MNRAS.457..232C}, an accreting NS-WD binary with strong dipole magnetic fields \citep[e.g.][]{2016ApJ...823L..28G,2020MNRAS.497.1543G,2022ApJ...939...27C}, the BH-NS or NS-NS mergers \citep[e.g.][]{2015ApJ...814L..20M,2016ApJ...822L...7W}, a pulsar passing through the asteroid belt \citep[e.g.][]{2016ApJ...829...27D,2020ApJ...895L...1D}, the collapse of a rotating supermassive NS to a BH \citep[e.g.][]{2014A&A...562A.137F,2014ApJ...780L..21Z}, and the mergers of charged BHs \citep[e.g.][]{2016ApJ...826...82L,2016ApJ...827L..31Z}. The multiband radiations and multimessenger signals are inevitable in these models.
	
The concept of tidal disruption events (TDEs)  originated in the late 1970s \citep[e.g.][]{1975Natur.254..295H}. In TDEs, the central objects are generally compact objects. When the tidal star passes the tidal radius of the central object, the tidal star is either tidally disrupted or swallowed by the central object, depending on the ratio of the tidal disruption radius to the pericenter distance \citep[e.g.][]{2021ARA&A..59...21G}. The BH-TDEs can produce the bright electromagnetic radiation flares visible from radio to X-ray even $\gamma$-ray \citep[e.g.][]{2008ApJ...684.1330L,2009ApJ...695..404R,2016ApJ...819....3M,2021ARA&A..59...21G}. Previous studies have shown that in TDEs, tidal stars are disintegrated by BH into debris streams, some of which return to the BH to form accretion streams that provide energy for the luminous mechanism of tidal events \citep[e.g.][]{1988Natur.333..523R}. An accretion disc formed by a BH with a high spin generates a magnetic field through the Blandford-Znajek (BZ) mechanism \citep{1977MNRAS.179..433B}, and the BZ jet power and the magnetic field strength of the disc are related to the accretion rate \citep[e.g.][]{2017NewAR..79....1L,2018pgrb.book.....Z}. Here we propose that a WD is tidally disintegrated by an intermediate-mass BH (IMBH). If a mass of WD fragments falling into the IMBH could cut the magnetic fields of the accretion disc, which might be trigger a RFRB.
	
In the following, we present the IMBH-WD TDE model and show the related properties of RFRBs in Section 2. The conclusions and discussion are made in Section 3.

\begin{figure*}
\centering
\includegraphics[width=0.5\textheight]{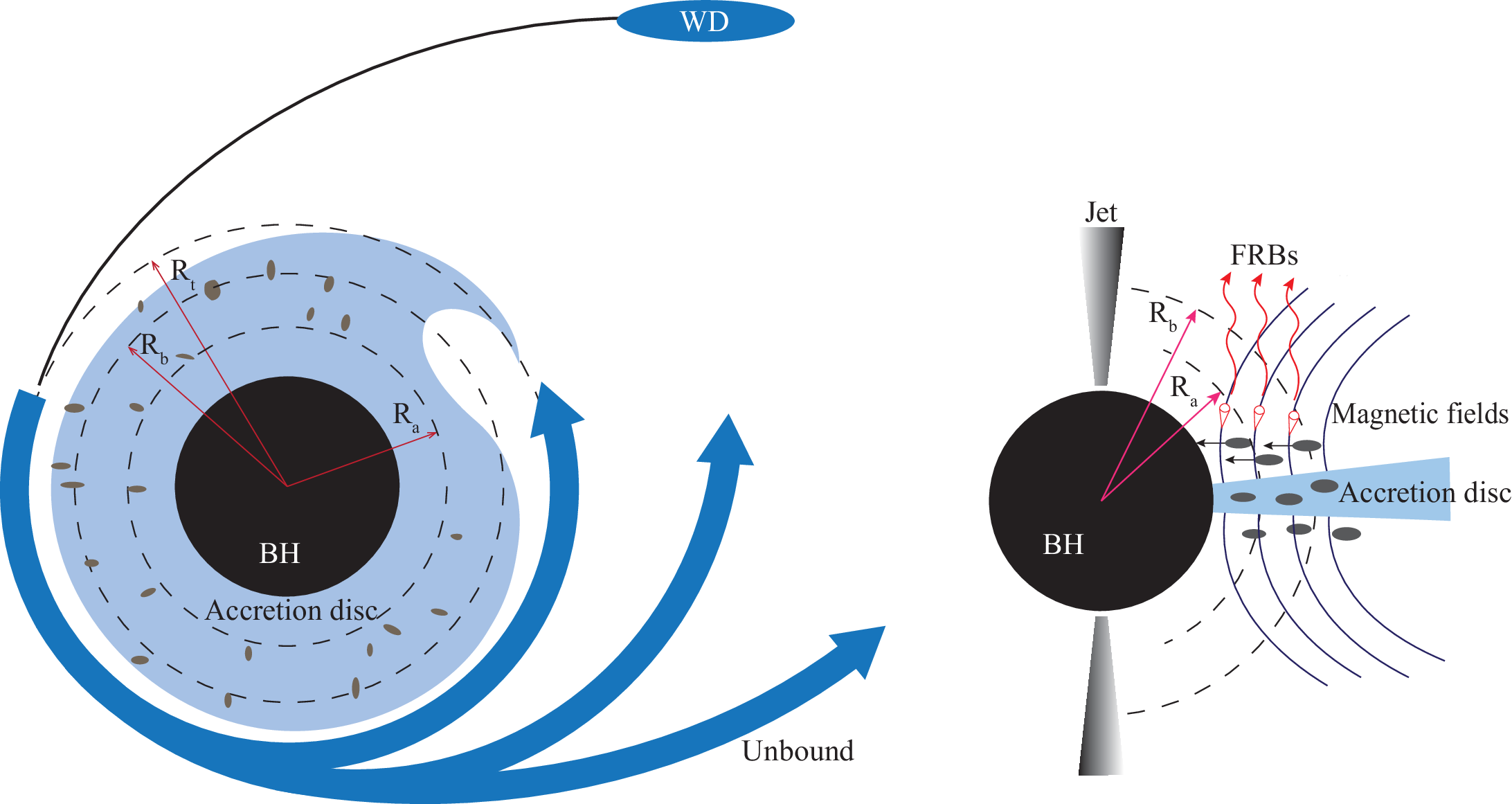}
\caption{Schematic diagram of tidal capture model. (i) A typical WD is tidally captured and destroyed by an IMBH at the tidal collapse radius $R_\mathrm{t}$, with a portion of the destroyed material escaping the BH at high speeds, and the rest being accreted by the BH and undergoing orbital circularization. (ii) The WD fragments retreat to the destruction radius $R_\mathrm{b}$ and are warped by tidal forces and destroyed again. (iii) The broken fragments cut the field lines of the accretion disc at the interaction radius $R_\mathrm{a}$, inducing RFRBs through coherent curvature radiation.}
\end{figure*}
	
\section{Model} \label{sec:THE CAPTURE MODEL}
	
We consider that as the WD approaches the tidal radius of an IMBH, it is captured and destroyed by the BH's powerful tidal forces into a debris stream, some of which is accreted to form an accretion disc and eventually falls into the BH, while other debris flies away from the edge of the BH with speeds $\ge 10^4 ~\rm km~s^{-1}$ \citep[e.g.][]{1988Natur.333..523R,2009ApJ...695..404R}. The formation of a transient accretion disc has such a high accretion rate that the inner region of the disc is dominated by radiation pressure. The accretion materials should inherit the magnetic fields of WDs, then a mass of WD fragments cutting the magnetic fields of the disc can provide energy of RFRBs.

\subsection{WD-IMBH TDEs}\label{sec:Tidal disruption event}

We set a typical WD with a mass of $M_\mathrm{wd}=0.5~M_{\mathrm{\odot}}$ and a radius of $R_\mathrm{wd}=1.1\times10^{4}~{\rm km}$, which has a tidal radius of \citep[e.g.][]{1988Natur.333..523R}
\begin{equation}
R_{\mathrm{t}}\approx R_{\mathrm{wd}}\left(\frac{M_\mathrm{BH}}{M_\mathrm{wd}}\right) ^{\frac{1}{3}}=1.39 \times 10^{10}{M_\mathrm{3}}^{\frac{1}{3}}~{\rm cm},
\label{eq1}
\end{equation}
where $M_3=M_\mathrm{BH}/(10^3~M_\odot)$. Then, a part of the broken part is tidally captured by the BH, and another part will be ejected at high speed, as shown in Figure 1. The captured part will form a transient accretion disc. For an IMBH, the Schwarzschild radius is $r_\mathrm{s}=2GM_\mathrm{BH}/c^{2}\sim2.96\times 10^{8}M_\mathrm{3}~\rm cm$, the event horizon radius is expressed as $r_\mathrm{h}=r_\mathrm{s}[1+(1-a_\mathrm{*}^{2})^{1/2}]/4 \sim 1.06\times 10^{8}M_\mathrm{3}~\rm cm$, and the spin
dimensionless constant $a_\mathrm{*}=0.9$ is taken in our model.
	
According to \cite{2008ApJ...679.1385R}, for the tidal collapse model of WD and IMBH, the initial material enters a circular orbit and forms an accretion surface, and the time for the initial material to return to the BH is $\le t_\mathrm{a}\sim 180(a/340r_\mathrm{s})^{3/2}M_\mathrm{3}^{-1/2}~\rm s$, where $a\sim 340 r_\mathrm{s} M_\mathrm{3}^{-1/3} r_\mathrm{wd,9} M_\mathrm{wd,0.5}^{-2/3}$ is the semi-major axis of the orbit, and $r_\mathrm{wd,9}= R_\mathrm{wd}/(10^9 ~\rm cm)$ and $M_\mathrm{wd,0.5}=M_\mathrm{wd}/(0.5~M_\odot)$. The viscous effect then releases enough binding energy to power the light-emitting mechanism. Over time, the vicinity of the BH will be replenished only by the injection of matter, and the accretion rate will gradually evolve, the evolution of the accretion rate over time is given by \citep{1988Natur.333..523R}
\begin{equation}
\dot{M}=7.94\times10^{-4} \left( \frac{M_\mathrm{BH}}{M_\mathrm{\odot}}\right) ^{-\frac{1}{2}} \left( \frac{t}{t_\mathrm{D}}\right) ^{-\frac{5}{3}} ~{M_\mathrm{\odot}~\rm s^{-1}},
\label{eq2}
\end{equation}
where $t_\mathrm{D}$ is the dynamic or orbital time scale given by \citep{1984A&A...131..267S}
\begin{equation}
t_\mathrm{D}=\left( \frac{GM_\mathrm{BH}}{R_\mathrm{t}^{3}}\right) ^{-\frac{1}{2}}=4.47 ~{\rm s}.
\label{eq3}
\end{equation}

We consider that the magnetic field is frozen in the disc plasma and transmitted to the BH through the inward motion of the plasma. The dynamics on the disc affects the value of the magnetic field, and the plasma with low magnetic pressure will be swallowed by the BH, while the plasma with high magnetic pressure will be pushed back to the disc and continue to be modulated. Only when the magnetic pressure of the disc and the pressure of the inner region of the disc are balanced, the magnetic field can maintain the strength suitable for the inner region of the disc. At the inner radius, the velocity of the plasma falling into the BH $v_r \sim c$. One can use this relation to estimate the magnetic strength \citep[e.g.][]{1997MNRAS.292..887G,1999ASPC..190..173L,2017NewAR..79....1L}, then we obtain
\begin{equation}
\frac{B^{2}}{8\pi}=P_\mathrm{in}\sim \rho_\mathrm{in}c^{2} \sim \dot{M} \frac{c}{4\pi r_\mathrm{ms}^{2}},
\label{eq4}
\end{equation}
where $r_\mathrm{ms}$ is the innermost stable orbit of the accretion disc, expressed as $r_\mathrm{ms}=r_\mathrm{s}[3+Z_{2}-\sqrt{(3-Z_{1})(3+Z_{1}+2 Z_{2})}]/2 \sim3.44\times 10^{8}M_\mathrm{3}~\rm cm$	with $Z_{1}=1+(1-a_{*}^{2})^{1/3}[(1+a_{*})^{1/3}+(1-a_{*})^{1/3}]$ and $Z_{2}=\sqrt{3 a_{*}^{2}+Z_{1}^{2}}$ \citep[e.g.,][]{1972ApJ...178..347B,1998GrCo....4S.135N}.
	
To simplify the calculation, we assume that the magnetic fields of the disc change with the radius, then the accretion rate can be substituted into the above formula to obtain the magnetic field changes with the radius and time, i.e.,
\begin{equation}
B=1.91 \times 10^{11}R_\mathrm{9}^{-1}M_\mathrm{3}^{-\frac{1}{4}}t_\mathrm{0}^{-\frac{5}{6}}~{\rm G},
\label{eq5}
\end{equation}
where $R_9=R/(10^9~\rm cm)$ is the dimensionless distance of the fragment to the BH and $t_\mathrm{0}= t/(10^{0}~\rm s)$. From the above formula, we can see that the evolution of the magnetic fields follows $t^{-5/6}$ in the accretion process.
	
\subsection{RFRBs}\label{R-FRBs}
	
We set a WD fragment with radius $r_\mathrm{0}$, mass $m_\mathrm{0}$, and density $\rho_\mathrm{0}$. When the fragment moves towards the center of the BH as shown in Figure 1, it will be tidally distorted by the BH at the breaking radius $R_\mathrm{b}$. The radius of breakage is given by \citep{1981ApJ...248..771C}
\begin{equation}
R_\mathrm{b}=\left( \frac{\rho_\mathrm{0} r_\mathrm{0}^{2} GM_\mathrm{BH}}{s}\right) ^{\frac{1}{3}}=9.45\times10^{9}m_\mathrm{27}^{\frac{2}{9}}\rho_\mathrm{7}^{\frac{1}{9}}s_\mathrm{19}^{-\frac{1}{3}}M_\mathrm{3}^{\frac{1}{3}}~{\rm cm},
\label{eq6}
\end{equation}	
where $m_{27}=m_0 /(10^{27} ~{\rm g})$ and $\rho_7=\rho_\mathrm{0}/ (10^7{\rm g~cm^{-3}})$. We set $10^{27}$ g as the typical WD fragment mass because for a larger one it is already shredded into smaller clumps before reaching the tidal radius. We take $s_\mathrm{19} =s/(10^{19}~\rm dyn~cm^{-2})$ in our model, and its typical value is set between the estimated tensile strength of different types of WDs \citep[e.g.][]{2022PhRvB.105b4110S}. The fragment mass, density, and tensile strength have their own parameter spaces and they are also related to each other. Here we just set the typical values to estimate the timescale and luminosity of FRBs. Although the typical values of these parameters are not completely certain that they will be produced in an IMBH-WD-TDE, the clumps between $\sim 10^{24}$ and $\sim 10^{28}~\rm g$ are likely to produce FRBs in our model. The broken fragments will be lengthwise stretched into a cylindrical-like shape with radius $r_\mathrm{0}$ and length $l_\mathrm{0}=2r_\mathrm{0}$  \citep[e.g.][]{1981ApJ...248..771C,2016ApJ...829...27D}. After that, it continuously broken into a series of small fragments as they approach the event horizon of the BH, where the time difference between the front and rear of fragments falling into the BH can be estimated as \citep[e.g.][]{2016ApJ...829...27D}	
\begin{equation}
\Delta t\approx\frac{12 r_\mathrm{0}}{5}\left( \frac{2GM_\mathrm{BH}}{R_\mathrm{b}}\right) ^{-\frac{1}{2}}=1.14m_\mathrm{27}^{\frac{4}{9}}\rho_\mathrm{7}^{-\frac{5}{18}}s_\mathrm{19}^{-\frac{1}{6}}M_\mathrm{3}^{-\frac{1}{3}}~{\rm ms}.
\label{eq7}
\end{equation}
It can be noted that this time difference is not only independent of the radius of the star, but also has a weak correlation with other parameters, and is numerically consistent with the typical duration of an FRB. The magnetic fields of the accretion disc must be affected during the falling of the debris.

The surface of the debris after tidal distortion will generate a strong induced electric field $\textbf{E}=-\textbf{v}_\mathrm{ff}\times \textbf{B}/c$, where $v_\mathrm{ff}=\left( {2GM_\mathrm{BH}}/{R} \right) ^{1/2}$ is the velocity of the debris in free fall under the gravity of the BH. When the magnetic field of the accretion disc is so strong that the magnetic pressure dominates the pressure in the inner region, the material at the edge of the BH's event horizon will quickly fall into the central BH in free fall \citep[e.g.][]{1974MNRAS.168..603L,2001MNRAS.320..235C}. We take the free fall time $t_\mathrm{ff}\sim R_\mathrm{b}/v_\mathrm{bf}$ describing the variability of accretion rate, and set $v_\mathrm{bf}=\left( {2GM_\mathrm{BH}}/{R_\mathrm{b}} \right) ^{1/2}$, which is used to calculate the magnetic field of the debris passing through the inner region of the accretion disc. It is worth mentioning that this timescale is the largest one for generating FRBs in our model. During the timescale of $t_\mathrm{ff}$, the values of the magnetic and electric fields in the inner region of the accretion disc can be given due to changes in the rate of the accretion disc, i.e.,
\begin{equation}
B=1.18\times10^{11}m_\mathrm{27}^{-\frac{5}{18}}\rho_\mathrm{7}^{-\frac{5}{36}}s_\mathrm{19}^{\frac{5}{12}}R_\mathrm{9}^{-1}M_\mathrm{3}^{-\frac{1}{4}}~{\rm G},
\label{eq8}
\end{equation}
and
\begin{equation}
E=6.44 \times 10^{10} m_\mathrm{27}^{-\frac{5}{18}}\rho_\mathrm{7}^{-\frac{5}{36}}s_\mathrm{19}^{\frac{5}{12}}R_\mathrm{9}^{-\frac{3}{2}}{M_\mathrm{3}}^{\frac{1}{4}}~{\rm volt~cm^{-1}}.
\label{eq9}	
\end{equation}
	
Here we adopt the method in \citet{2016ApJ...829...27D}, then assume that the magnetic energy density obtained by the magnetic dipole moment $\mu \sim B(R_\mathrm{b})R_\mathrm{b}^3$ at $R_b$ is equal to the kinetic energy density $\rho v_\mathrm{af}^{2}/2$ at the interaction radius, i.e.,	
\begin{equation}
R_\mathrm{a}=\left( \frac{2\mu^{2}}{8\pi \rho v_\mathrm{af}^{2}}\right) ^{\frac{1}{6}}=3.19\times10^{8} m_\mathrm{27}^{\frac{1}{	15}}\rho_\mathrm{7}^{-\frac{1}{6}}s_\mathrm{19}^{-\frac{1}{10}}M_\mathrm{3}^{-\frac{1}{30}}~{\rm cm},
\label{eq10}
\end{equation}
where $v_\mathrm{af}=\sqrt{2GM_\mathrm{BH}/R_\mathrm{a}}$. It should be noted that $R_\mathrm{a}$ is obtained under the assumption of magnetic moment $\mu$, and $r_\mathrm{h}<R_\mathrm{a}<R_\mathrm{b}$ is necessary so that the effect of the magnetic field on the fall of the fragments is negligible. In addition, $R_\mathrm{a}$ is close to the event horizon radius, so the time it takes for the debris to cross $R_\mathrm{a}$ and release the FRB can be approximated to $\Delta t$. These results make our model self-consistent.

Because the electric field on the surface of the fragment has a parallel component to the magnetic field, the electrons are accelerated to the relativistic energies and torn apart from the fragment's surface \citep[e.g.][]{1991PhFlB...3.1892G,1992JGR....9712057N}. Motion along a magnetic inductance line of curvature $\rho_\mathrm{c}$ generates coherent curvature radiation \citep[e.g.][]{1975ApJ...196...51R}, and the maximum Lorentz factor of the accelerating electron can be obtained by assuming that the synchrotron radiation power $\sigma_\mathrm{T}c\gamma_\mathrm{max}^{2}B^{2}/6\pi$ and the accelerating power of the electron $eEc$ are equal \citep[e.g.][]{2016ApJ...829...27D,2020ApJ...897L..40D}, thus
\begin{equation}
\gamma_\mathrm{max}=\chi\left( \frac{6\pi eE}{\sigma_\mathrm{T}B^{2}}\right) ^{\frac{1}{2}}=250.32 \chi m_\mathrm{27}^{\frac{5}{36}}\rho_\mathrm{7}^{\frac{5}{72}}s_\mathrm{19}^{-\frac{5}{24}}R_\mathrm{9}^{\frac{1}{4}}{M_\mathrm{3}}^{\frac{3}{8}},
\label{eq11}
\end{equation}
where $\sigma_\mathrm{T}$ is the Thomson scattering cross section. $\chi$ is defined as the proportional coefficient between $\gamma$ and the maximum Lorentz factor \citep[e.g.][]{2020ApJ...897L..40D}, and the conditions need to be met $\chi \le 1$. We assume that $\chi \sim 0.2$, substitute it into Equation (\ref{eq12}) to get $\gamma \sim 50.06$.
	
We note that the breaking radius $R_\mathrm{b}\sim 9.45 \times 10^{9}~{\rm cm}$ is less than the tidal radius $R_\mathrm{t}\sim 1.39 \times 10^{10}~{\rm cm}$ and greater than the interaction radius $R_\mathrm{a}\sim 3.19 \times 10^{8}~{\rm cm}$, which means that our model is self-consistent. Next, we discuss the radiative properties of coherent radiation. For a ultrarelativistic electron moving along a magnetic inductance line with radius of curvature $\rho_\mathrm{c}$ and Lorentz factor $\gamma$, the characteristic frequency is given by \citep[e.g.,][]{2016ApJ...829...27D}
\begin{equation}
\nu_\mathrm{curv}=\frac{3c\gamma^{3}}{4\pi\rho_\mathrm{c}}=4.43 m_\mathrm{27}^{\frac{7}{20}}\rho_\mathrm{7}^{\frac{3}{8}}s_\mathrm{19}^{-\frac{21}{40}}R_\mathrm{9}^{\frac{3}{4}}M_\mathrm{3}^{\frac{139}{120}}~{\rm GHz},
\label{eq12}
\end{equation}
where the radius of curvature is $\rho_\mathrm{c}\sim0.635R_\mathrm{a}$ given by \cite{2018ApJ...868...31Y}, which is consistent with the characteristic frequency of typical FRBs. In theory, the WD fragments would be laterally compressed into a series of smaller fragments after the breaking radius and continue to move along the closed magnetic field lines. The ultrarelativistic electrons carried by the debris produce the coherent curvature radiation, emitting a bright radio burst, and forming a sector with an angle of $\theta=r (R_\mathrm{a}) /R_\mathrm{a}$. We consider a region surrounded by a circuit and divide the region into slices of coherent radiation, the luminosity of the coherent curvature radiation is given by \citep[e.g.,][]{2016ApJ...829...27D}
\begin{equation}
\begin{aligned}
L_\mathrm{tot}&\approx P_\mathrm{c}N_\mathrm{e,tot}^{2}N_\mathrm{slice}^{-1}\\
&\approx1.96\times10^{40}m_\mathrm{27}^{\frac{17}{20}}\rho_\mathrm{7}^{-\frac{11}{8}}s_\mathrm{19}^{\frac{29}{40}}R_\mathrm{9}^{-\frac{19}{4}}{M_\mathrm{3}}^{\frac{229}{120}}~{\rm{erg~s^{-1}}},
\label{eq13}
\end{aligned}
\end{equation}
where $P_\mathrm{c}$ is the curvature radiation power of a ultrarelativistic electron, $N_\mathrm{e,tot}$ is the total number of electrons in the emission region, and $N_\mathrm{slice}$ is the number of coherent slices. It can be seen from the formula that the total luminosity $L_\mathrm{tot}$ has a strong correlation with $R$.

According to the law of conservation of energy, we know that the coherent curvature radiation discussed above is actually provided by the gravitational binding energy of the BH. The total released gravitational energy, $E_G\sim 10^{47}~\rm erg$, is definitely higher than the energy of FRBs. From Equation (\ref{eq5}) we can see that $B\propto t^{-\frac{5}{6}}$, for our selection of $t_\mathrm{ff}$, we get the minimum value of the magnetic field, this means that Equation (\ref{eq13}) is the minimum luminosity limit for our model to produce FRBs. For the existing observations, this power meets the luminosity range of the FRB, which also indicates that in addition to the fragment parameters selected here, with the change of the distance between the fragment and the BH event horizon, the change of the size of the tidal fragment, and the evolution and configuration of the magnetic fields also has the possibility of producing FRB in different directions.

In addition, the RFRBs induced by the IMBH-WD TDE might be last in the whole accretion process, especially the early stage with high accretion rate and consequentially strong magnetic fields.

\subsection{Event rate estimation}

In our model, we believe that if FRBs are to be generated in TDEs, then the mass range of the BHs should be satisfied $\sim 10^3-10^5 M_\mathrm{\odot}$. Once the BH is too massive $(\ge 10^5 M_\mathrm{\odot})$, the WD is swallowed whole instead of being tidally destroyed. Assuming that the total number of IMBHs in the galaxy is $N$, the proportion of BHs in the mass range is $k$, and the rate of TDEs between IMBHs and WDs is $\Gamma_\mathrm{t}$, then we can figure out the probability of TDEs in the galaxy \citep[e.g.][]{2008ApJ...684.1330L}, i.e., $\Gamma_\mathrm{TDE}=k\Gamma_\mathrm{t}N$. We use assumption in \cite{2016ApJ...819....3M} that the IMBH mass function is flat with a lower mass $(M_\mathrm{BH}< 10^6 M_\mathrm{\odot})$, the event rate can be estimate as
\begin{equation}
\Gamma_\mathrm{TDE}=k\Gamma_\mathrm{t}N \sim 10{\dot{N}}_\mathrm{IMBH,-6}n_\mathrm{IMBH,7}~\rm {yr^{-1}~Gpc^{-3}},
\label{eq14}
\end{equation}
where ${\dot{N}}_\mathrm{IMBH,-6}={\dot{N}}_\mathrm{IMBH}/(10^{-6} ~\rm yr^{-1})$ and $n_\mathrm{IMBH,7}=n_\mathrm{IMBH}/(10^7 ~\rm Gpc^{-3})$. This rate can be roughly considered as the FRB event rate.

Observational multi-band radiations of TDEs are generally considered to be emitted from the accretion discs and only a few fraction is driven by the jets \citep[e.g.][]{2011ApJ...740L..27L,2013ApJ...762...98L,2016ApJ...816...20L,2020SSRv..216...81A}. In our model, FRBs might be radiated along the magnetic field lines as shown in Figure 1, so one can be cautiously optimistic for the discovery on IMBH-WD TDEs associated with FRBs.

\section{Conclusions and discussion} \label{sec:conclusions}

In this \emph{Letter}, we present a model of the tidal disintegration of WD by IMBH for a unified explanation of the origin of extragalactic RFRBs. The main conclusions are as follows:
\begin{enumerate}
\item
A WD ($\sim 0.5~M_\odot$) is tidally captured by the IMBH ($\sim 10^3~M_\odot$) at the tidal radius $R_\mathrm{t}\sim 1.39 \times 10^{10}~{\rm cm}$, during the capture process, a transient accretion disc is formed around the BH, and the accretion rate changes with time, $\dot{M}\propto t^{-5/3}$. The magnetic field strength near the BH event horizon is large enough to trigger FRBs.
		
\item
Due to the tidal shear of the IMBH, the WD fragments after the tidal collapse in the accretion disc are again distorted at the breaking radius $R_\mathrm{b}\sim 9.45 \times 10^{9}~{\rm cm}$ into a series of smaller fragments. The time difference between the leading and trailing fragments of this series of fragments reaches the edge of the BH's event horizon is $\Delta t\sim1.14~{\rm ms}$, which is consistent with the typical pulse duration of RFRBs.
		
\item
Within an emission region surrounded by a circuit, the surface electrons of the fragments are accelerated to the ultrarelativistic energy by the strong magnetic field, and are torn from the surface to move along the magnetic inductance line, resulting in curvature radiation. The electrons carried by a series of fragments produce coherent curvature radiation with a characteristic frequency of $\mu_\mathrm{curv}\sim 4.43~{\rm GHz}$. The luminosity can be estimated to be $L_\mathrm{tot}\sim 1.96\times10^{40}~{\rm erg~s^{-1}}$, which meets the typical luminosity of FRBs.

\item
RFRBs induced by IMBH-WD TDEs are short-lived, multiple, and irregular transients throughout the accretion processes. Their event rate might be $\sim 10~\rm {yr^{-1}~Gpc^{-3}}$.
\end{enumerate}

Actually, besides IMBH-WD TDEs, the trigger sources of fast radio transients, i.e., high-velocity compact fragments cutting magnetic lines, exist in some multibody systems, such as the magnetars surrounded by the unstable asteroid belts \citep[e.g.][]{2016ApJ...829...27D,2023MNRAS.523.2732W}. The IMBH-WD TDEs are usually accompanied by the generation of high-energy transients, and are also the main goals of the space-based gravitational wave projects, such as LISA, TianQin, and Taiji. The future multiband and multimessenger detections might discover the RFRBs induced by TDEs.

In TDEs, except the parts accreted by BH, those clumps that managed to escape or were not swallowed by the BH are also of interest, and there has been some related work discussed the ultimate fate of these clumps \citep[e.g.][]{2020ApJ...904..100R,2023MNRAS.524.3026B}. Our future work will consider which clumps are trapped by BH but not falling into BH to form compact planets, as well as the possibility that ejected clumps to form high-speed wandering planets.
	
\section*{Acknowledgements}

We thank anonymous referee for helpful suggestions that improved the manuscript, and also thank Wei-Min Gu, De-Fu Bu, Ang Li, Wei-Hua Lei, Xiao-Yan Li, Bao-Quan Huang, and Ali Shad for helpful discussion. This work was supported by the National Natural Science Foundation of China under grants 12173031 and 12221003.

\section*{Data availability}

The data underlying this article will be shared on reasonable request to the first author.

\clearpage
\end{document}